\documentclass[12pt]{iopart}
\usepackage{iopams}
\usepackage{epsfig,graphics,color}
\newcommand \etc {{\it etc.} }
\newcommand \tie {{\it i.e.}}

\newcommand \kd  {\delta}
\newcommand \ra  {\rightarrow}

\newcommand \vk {\vec{k}}

\newcommand \vx {\vec{x}}

\newcommand \vecr {\vec{r}}

\newcommand \g {\gamma}

\newcommand \ep {\epsilon}

\newcommand \p {^{\prime}}

\newcommand \x {\cdot}
\newcommand \hf {\frac{1}{2}}
\newcommand \A {\alpha}

\newcommand \lc {\langle}
\newcommand \rc {\rangle}
\newcommand \prt {\partial}
\newcommand \D {\Delta}
\newcommand \sg {\sigma}

\newcommand \nt {\noindent}

\newcommand {\llb} { \left[ \frac{\mbox{}}{\mbox{}} \right.}
\newcommand {\lrb} { \left. \frac{\mbox{}}{\mbox{}} \right] }

\newcommand \bvec{\left( \begin{array}{c} }
\newcommand \evec{\end{array} \right)}

\newcommand \eg {{\it e.g.}}
\newcommand \bea{\begin{eqnarray} }
\newcommand \eea{\end{eqnarray} }

\newcommand {\be} {\begin{equation}}
\newcommand {\ee} {\end{equation}}

\newcommand {\gev} {\mbox{GeV}}

\newcommand{\psibar} {\bar{\psi}}

\begin{document}

\title[Jet Quenching]{A comparative study of Jet-quenching Schemes}

\author{A.~Majumder}

\address{Department of Physics, Duke University, Durham NC 27708.}
% \ead{abhijit.majumder@duke.edu}
\begin{abstract}
The four major approximation schemes devised to study the modification of jets 
in dense matter are outlined. The comparisons are restricted to basic assumptions 
and approximations made in each case and the calculation methodology used. 
Emergent underlying similarities between apparently disparate methods 
brought about by the approximation schemes are exposed. 
Parameterizations of the medium in each scheme are discussed in terms of the 
transport coefficient $\hat{q}$.  Discrepancies between the estimates 
obtained from the four schemes are discussed.  Recent developments in the 
basic theory and phenomenology of energy loss are highlighted.
\end{abstract}

%\maketitle

%%%%%%%%%%%%%%%%%%%%%%%%%%%%%%%%%%%%
%%%%%%%%%%%%%%%%%%%%%%%%%%%%%%%%%%%%
%%%%%%%%%%%%%%%%%%%%%%%%%%%%%%%%%%%%

\section{Introduction}

%%%%%%%%%%%%%%%%%%%%%%%%%%%%%%%%%%%%
%%%%%%%%%%%%%%%%%%%%%%%%%%%%%%%%%%%%
%%%%%%%%%%%%%%%%%%%%%%%%%%%%%%%%%%%%

The considerable modification of  the spectrum of high transverse momentum $(p_T)$ hadrons 
produced in ultra-high energy heavy-ion collisions has now been established by experiments 
at the Relativistic Heavy-Ion Collider (RHIC). 
The number of such hadrons, irrespective of flavour,  with $p_T \geq 7$GeV 
is reduced by almost a factor of 5 in central $Au$-$Au$ collisions compared to that 
expected from elementary nucleon nucleon encounters 
enhanced by the number of expected binary collisions~\cite{Adcox:2001jp,Adler:2002xw}. 
This suppression, referred to as \emph{jet-quenching}, 
caused by the energy loss of  very energetic partons produced in 
the few  hard  scatterings  that occur early in the collision, represents one of the 
major theoretical predictions which have been substantiated by experiment~\cite{Wang:1991xy,Baier:1996kr}.
The basic underlying mechanism, that of induced gluon radiation from the hard parton traversing a 
coloured environment, has by now been 
expounded upon in multiple reviews~\cite{Baier:2000mf,Gyulassy:2003mc} and 
there exists a certain consensus regarding the physics involved. 

In the last several years, dramatic progress has been made both in experiment and theory. While measurements 
on single inclusive observables have been extended to wider ranges in $p_T$ and over a variety of systems, there 
has arisen a plethora of  multiparticle jet-like correlation observables, photon-jet, jet-medium and heavy flavour 
observables~\cite{RHIC_Whitepapers}\footnote{See also Ref.~\cite{QM2006_talks} and references therein}.
On the theory side, calculations have evolved through multiple levels of complexity and have also been expanded 
in scope to address the wide variety of new observables. 
In the following, the status of energy loss schemes in the light flavour sector are reviewed.  New 
observables capable of offering deeper insight into the structure of the dense matter produced 
and differentiating between the various schemes, as well as new developments in the theory of jet modification and 
correlations are outlined. 

%%%%%%%%%%%%%%%%%%%%%%%%%%%%%%%%%%%%
%%%%%%%%%%%%%%%%%%%%%%%%%%%%%%%%%%%%
%%%%%%%%%%%%%%%%%%%%%%%%%%%%%%%%%%%%

\section{The four schemes}

%%%%%%%%%%%%%%%%%%%%%%%%%%%%%%%%%%%%
%%%%%%%%%%%%%%%%%%%%%%%%%%%%%%%%%%%%
%%%%%%%%%%%%%%%%%%%%%%%%%%%%%%%%%%%%

\begin{figure}
\hspace{1cm}
\resizebox{2in}{2in}{\includegraphics[0in,0in][5in,5in]{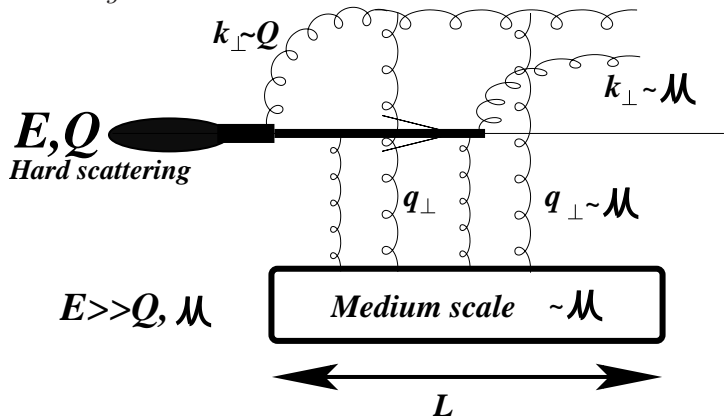}}
\caption{A schematic picture of the various scales involved in the modification of jets in dense matter. }
\label{fig0}
\end{figure}

The majority of current approaches to the energy loss of light partons may 
be divided into four major schemes often referred to by the names of the original 
authors.
All schemes utilize a factorized approach where the 
final cross section to produce a hadron $h$  with transverse momentum $p_T$ 
(rapidity between $y$ and $y+dy$) 
may be expressed as a convolution of initial nuclear structure functions [$G_a^A(x_a),G_b^B(x_b) $, initial state 
nuclear effects such as shadowing and Cronin effect are understood to be included] to produce 
partons with momentum fractions $x_a,x_b$, a 
hard partonic cross section to produce a high transverse momentum parton $c$ with a 
transverse momentum $\hat{p}$ and a medium 
modified fragmentation function for the final hadron [$\tilde{D}_c^h(z)$], 

\bea
\fl \frac{d^2 \sg^h}{dy d^2 p_T} = \frac{1}{\pi} \int dx_a \int d x_b G^A_a(x_a) G^B_b(x_b) 
\frac{d \sg_{ab \ra cX} }{d \hat{t}} \frac{\tilde{D}_c^h(z)}{z}. \label{basic_cross}
\eea

\nt
In the vicinity of mid-rapidity, $z=p_T/\hat{p}$ and $\hat{t} = (\hat{p} - x_a P)^2$  ($P$ is the 
average incoming momentum of a nucleon in nucleus A).
The entire effect of energy loss is concentrated in the  calculation of the 
modification to the fragmentation function. The four models of energy loss are in a  
sense four schemes to estimate this quantity from perturbative QCD calculations.  
To better appreciate the approximation schemes, one may introduce a set of scales (see Fig.~\ref{fig0}):
$E$ or $p^+$, the forward energy of the jet, $Q^2$, the virtuality of the initial jet-parton, $\mu$, 
the momentum scale of the medium and $L$, its spatial extent. Most of the   
differences between the various schemes may be reduced to the different 
relations between these various scales assumed by each scheme as well 
as by how each scheme treats or approximates the structure of the medium. In 
all schemes, the forward energy of the jet far exceeds the medium scale, $E >> \mu$. The schemes 
are presented from one extreme of the approximation set (higher twist approach)  to the opposite extreme 
(finite temperature approach), similarities in intermediate steps of the calculation will not be repeated.

%%%%%%%%%%%%%%%%%%%%%%%%%%%%%%%%%%%%
%%%%%%%%%%%%%%%%%%%%%%%%%%%%%%%%%%%%

\subsection{Higher Twist}

%%%%%%%%%%%%%%%%%%%%%%%%%%%%%%%%%%%%
%%%%%%%%%%%%%%%%%%%%%%%%%%%%%%%%%%%%

The origin of the higher twist (HT) approximation scheme lies in the calculations of 
medium enhanced higher twist corrections to the total cross section in  
Deep-Inelastic Scattering (DIS) off large nuclei~\cite{Qiu:1990xx}. The essential 
idea lies in the inclusion of power corrections to the leading twist cross sections, 
which, though suppressed by powers of the hard scale $Q^2$, are enhanced by the 
length of the medium. This technology of identifying and isolating power corrections 
is used to compute the single hadron inclusive cross-section. 

\begin{figure}
\hspace{1cm}
\resizebox{1.75in}{1.25in}{\includegraphics[0.5in,0.5in][5in,3.5in]{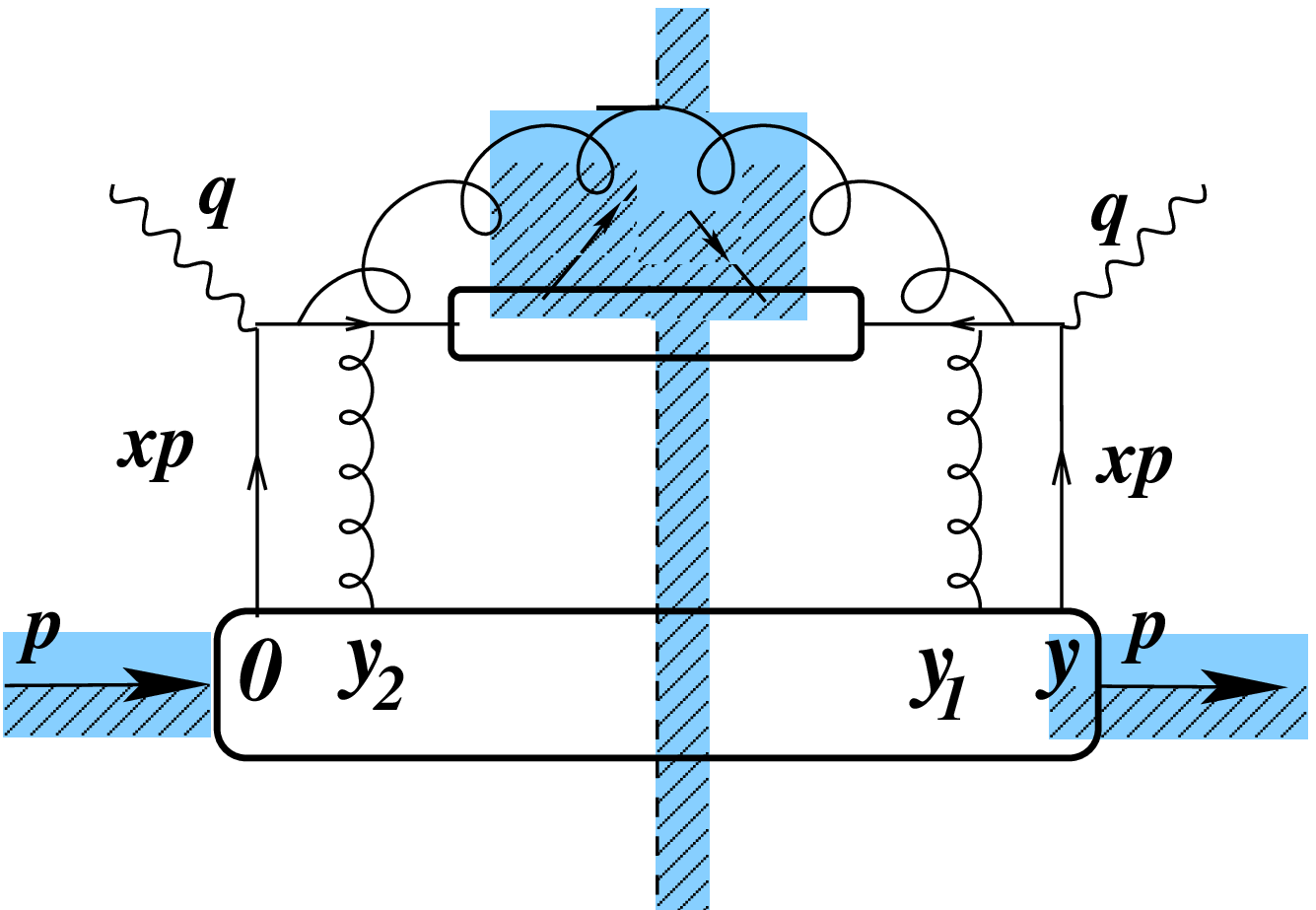}}
\hspace{1cm}
\resizebox{2.in}{1.25in}{\includegraphics[0.0in,-0.5in][7in,3.0in]{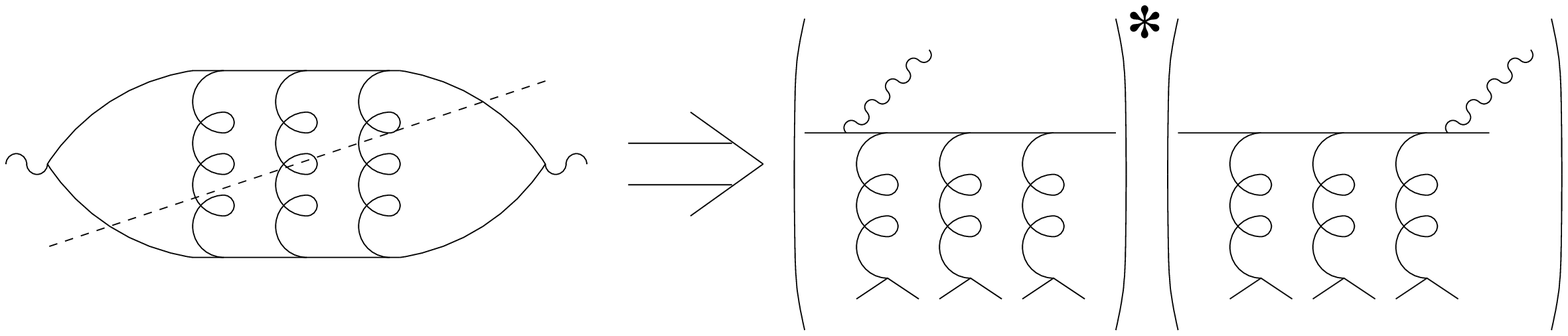}}
\caption{Left panel: a higher twist contribution to the modification of the fragmentation 
function in medium. Right panel: a typical cut diagram in the AMY formalism.}
\label{fig1}
\end{figure}

One assumes the 
hierarchy of scales $E >> Q >> \mu$ and applies this to the computation of multiple Feynman 
diagrams such as the one in the left panel of Fig.~\ref{fig1}, this diagram represents 
the process of a hard virtual quark produced in a hard collision, which then radiates 
a gluon and then scatters off a soft medium gluon with transverse momentum 
$q_\perp \sim \mu$ prior to exiting the medium and fragmenting into hadrons. 
At a given order, there exist various other contributions which involve scattering of 
the initial quark off the soft gluon field prior to radiation as well as scattering of the 
radiated gluon itself. All such contributions are combined coherently to calculate the 
modification to the fragmentation function directly.

The hierarchy of scales allows 
one to use the collinear approximation to factorize the fragmentation function and its 
modification from the hard scatterring cross section. 
Thus, even though such a modified 
 fragmentation function is derived in DIS, it may be generalized to the kinematics of a 
heavy-ion collision. 
Diagrams where the outgoing parton scatters off the medium gluons, 
 such as those in Fig.~\ref{fig1},
produce a medium dependent additive contribution to the vacuum fragmentation function, 
which may be expressed as, 

\bea
\fl \D D_i(z,Q^2) = \int_0^{Q^2} \frac{dk_{\perp}^2}{k_{\perp}^2} 
\frac{\A_s}{2\pi} \left[ \int_{z_h}^1 \frac{dx}{x} 
\sum_{j=q,g}\left\{ \D P_{i \ra j} (x,x_B,x_L,k_\perp^2) 
D_j^{h} \left(\frac{z_h}{x} \right) 
 \right\} \right].
 \label{med_mod}
\eea
\nt
In the above equation,  $\D P_{i\ra j}$ 
represents the medium modified splitting function of parton $i$ into $j$ 
where a momentum fraction $x$ is left in parton $j$. 
The factor, $x_L = k_\perp^2/(2P^-p^+ x(1-x))$ \footnote{Throughout these proceedings, four-vectors  
will often be referred to using the light cone convention where $x^\pm = (x^0 \pm x^3)/\sqrt{2}$. For the 
higher-twist scheme, often, $x^+ = ( x^0 + x^3 )/2$ and $x^- = x^0 - x^3$.}, where 
the radiated gluon or quark carries away a transverse momentum $k_\perp$, $P^-$ is 
the incoming momentum of a nucleon in the nucleus and $p$ is the momentum of the 
virtual photon. The medium modified splitting functions may be expressed as a product 
of the vacuum splitting function $P_{i \ra j}$ and a medium dependent factor, 

\bea
\fl \D \hat{P}_{i\ra j} &=& P_{i \ra j}  \frac{C_A 2\pi \A_s   T^A_{qg} (x_a,x_L)}{(k_\perp^2 + 
\lc q_\perp^2 \rc)  N_c f_q^A(x_a)} .  \label{mod_split}
\eea
\nt
Where, $C_A,N_c$ represent the adjoint Casimir and the number of colours. 
The mean 
transverse momentum of the soft gluons is represented by the factor $\lc q_\perp^2 \rc$.  
The term $T^A_{qg}$ represents the quark gluon correlation in the
nuclear medium, and depends on the four point correlator,  

\bea 
\fl \lc  P | \psibar (0)  \g^+ F_\sg^+ \!( y_2^- ) {F^{+}}^\sg\!(y_1^- ) \psi (y^-) | P \rc  
\!\sim\!  C \lc p_1| \psibar (0)  \g^+ \psi (y^-) | p_1 \rc 
 \lc p_2 |  F_\sg^+\!( y_2^- ) {F^{+}}^\sg \!(y_1^- )| p_2 \rc. \label{FF}
\eea
\nt
Where, $F_\sg^+ ( y_2^- )$ and $ {F^{+}}^\sg (y_1^- )$ represent gluon field 
operators at the locations $y_1^-,y_2^-$ and $\psi(y)$ represents the quark field operator. 
The above correlation function cannot be calculated from first principles without making 
assumptions regarding the structure of the medium. 
The only assumption 
made is that that the colour correlation length is small.
As a result, one may factorize 
the four point function into two separate structure functions, one for the original parton produced in the 
hard scattering [this is a quark in Eq.~\eref{FF}] and one for the soft gluon off which 
the parton scatters in the final state.

The entire phenomenology of the medium is incorporated as a model for the expectation of the 
second set of operators in Eq.~\eref{FF}. The product of its absolute magnitude as well as 
the correlation constant $C$ is set by fitting to one data point. Unlike the remaining formalisms, 
the H-T approach is set up to directly calculate the medium modified fragmentation function 
and as a result the final spectrum. 
The determined constant, $C$, may be used to calculate
the average energy loss encountered by a jet. Another advantage of this approach is the 
straightforward generalization to multiparticle correlations~\cite{Majumder:2004wh} and 
their modification in the medium.

%%%%%%%%%%%%%%%%%%%%%%%%%%%%%%%%%%%%
%%%%%%%%%%%%%%%%%%%%%%%%%%%%%%%%%%%%

\subsection{Opacity expansion: Reaction Operator Approach}

%%%%%%%%%%%%%%%%%%%%%%%%%%%%%%%%%%%%
%%%%%%%%%%%%%%%%%%%%%%%%%%%%%%%%%%%%

Unlike the higher twist or the finite temperature approaches, 
this formalism, often referred to as the Gyulassy-Levai-Vitev (GLV) scheme~\cite{Gyulassy:1999zd}, 
and other opacity expansion schemes were constructed primarily to deal with the 
problem of energy loss in dense deconfined matter. 
The GLV scheme assumes the medium to be composed of heavy almost static 
colour scattering 
centers which are well separated in the sense that the mean free path of a jet 
$\lambda \gg 1/\mu$ the colour screening length of the medium~\cite{Gyulassy:1993hr} . 
The opacity of the medium $\bar{n}$ quantifies the number of scattering centers seen  by a 
jet as it passes through the medium, \tie, $\bar{n} = L/\lambda$, where $L$ is the 
thickness of the medium. The opacity or gluon number density is the quantity used to 
model the presence of the medium.

A hard jet is produced locally in such a plasma with a large forward 
energy $p^+ \gg \mu$ and almost immediately begins to shower soft gluons. 
The typical transverse momentum of the radiated gluons is similar in order of magnitude to the 
transverse momentum imparted from the medium \tie, $E>> Q \sim \mu$.
The colour 
centers are assumed to produce a screened Yukawa potential. 
At first order in opacity, a jet scatters off 
one such potential and picks up a 
transverse momentum $\vec{q}_\perp$; in the process it will radiate a gluon with 
momentum $k \equiv ( xp^+,\frac{k_\perp^2}{xp^+}, \vk_\perp)$. The scattering 
may happen before or after the radiation. Squaring such contributions and including 
interference terms between vacuum radiation and double scattering, leads, in the 
limit of $x \ra 0$ (and ignoring spin effects), to the soft gluon differential emission 
distribution at first order in 
opacity~\cite{Gyulassy:1999zd}, 

\bea
\fl x\frac{dN}{dx dk_\perp^2} &=& x \frac{dN}{dx dk_\perp^2} \frac{L}{\lambda_g} 
\int_0^{q_{max}^2} d^2 {q}_\perp \frac{\mu_{eff}^2}{\pi ({q}_\perp^2 + \mu^2 )^2}
\frac{2 k_\perp \x {q}_\perp ( k- q_1)^2 L^2  }{ 16x^2 E^2 + (k - q)_\perp^2L^2}. \label{GLV2}
\eea
\nt
In the above equation, $\lambda_g$ is the mean free path of the radiated gluon. Consideration 
of single inclusive gluon emission from multiple scattering requires the use of a recursive diagrammatic procedure~\cite{Gyulassy:2000er}. The inclusion of such diagrams allows for the computation of 
gluon distributions to finite order ($n\geq 1$) in opacity.  

Due to the soft limit \tie, $x\ra 0$ used, multiple gluon emissions are required 
for a substantial amount of 
energy loss. Each such emission at a given opacity is assumed independent and a 
probabilistic scheme is set up, wherein, the jet loses an energy fraction $\ep$ in 
$n$ tries with a Poisson distribution~\cite{Gyulassy:2001nm}, 

\bea
\fl P_n(\ep,P^+) = \frac{e^{-\lc N_g \rc} }{n!} \Pi_{i=1}^n \llb \int d x_i \frac{dN_g}{dx_i} \lrb 
\kd(\ep - \sum_{i=1}^{n} x_i  ) , \label{GLV3}
\eea

\nt
where, $\lc N_g \rc$ is the mean number of gluons radiated per coherent interaction set.
Summing over $n$ gives the probability  $P(\ep)$ for  an incident jet to lose 
a momentum fraction $\ep$ 
due to its passage through the medium. This is then used to model a medium 
modified fragmentation function, by shifting the energy fraction available to 
produce a hadron (as well as accounting for the phase space available after energy loss), 

\bea
\fl \tilde{D}(z,Q^2) = \int_0^{1} d\ep  P(\ep) \frac{D\left( \frac{z}{1-\ep},Q^2\right)}{1-\ep}. \label{GLV4}
\eea
\nt
The above, modified fragmentation function is then used in a factorized formalism as in Eq.~\eref{basic_cross} 
to calculate the final hadronic spectrum.

%%%%%%%%%%%%%%%%%%%%%%%%%%%%%%%%%%%%
%%%%%%%%%%%%%%%%%%%%%%%%%%%%%%%%%%%%

\subsection{Opacity expansion: Path Integral Approach}

%%%%%%%%%%%%%%%%%%%%%%%%%%%%%%%%%%%%
%%%%%%%%%%%%%%%%%%%%%%%%%%%%%%%%%%%%

The path integral approach for the energy loss of  a hard jet propagating 
in a coloured medium was first introduced in Ref.~\cite{Zakharov:1996fv}.
It was later demonstrated to be equivalent to the well known BDMPS 
approach~\cite{BDMPS}. The current, most widespread, variant of this 
approach developed by numerous authors is often referred to as the 
Armesto-Salgado-Wiedemann (ASW) approach. In this scheme, one 
incorporates the effect of multiple scattering of the incoming  and 
outgoing partons in terms of a path integral over a path ordered 
Wilson line~\cite{Wiedemann:2000ez}.   

Similar to the GLV approach, this formalism also assumes a model for 
the medium as an assembly of Debye screened heavy scattering centers. 
A hard, almost on shell,  parton traversing such a medium will 
 engender multiple transverse scatterings of  order $\mu \ll p^+$. It will in the 
process split into an outgoing parton and a radiated gluon which will also scatter 
multiply in the medium. The radiation, being brought about by the multiple scattering 
has a transverse momentum $k_\perp \sim \mu$ (similar to the GLV and different from the HT approach).  
The propagation of the incoming (outgoing) partons as well as that of the radiated gluon 
in this background colour field may be expressed in terms of  effective Green's functions 
[$G(\vecr_\perp,z ; \vecr_{\perp}\,\p, z\p)$ (for quark or gluon)] which obey the obvious 
Dyson-Schwinger 
equation, 

\bea
\fl G(\vecr_\perp,z ; \vec{r\p}_\perp, z\p) = G_0(\vecr_\perp,z ; \vec{r\p}_{\perp}, z\p) -i \int_z^{z\p}\!\!\!\!\!\! d \zeta \!\!\int \!\!\! d^2\vx 
G_0(\vecr_\perp, z; \vx, \zeta) A_0(\vx,\zeta) G (\vx, \zeta; \vec{r\p}_{\perp}, z\p)  \label{ASW1},
\eea
\nt
where, $G_0$ is the free Green's function and $A_0$ represents the colour potential of  the medium. 
The solution for the above interacting Green's function involves a path ordered Wilson line  which 
follows the potential from the location $[\vecr_{\perp}(z\p), z\p]$ to $[\vecr_{\perp}(z), z]$. Expanding 
the expression for the radiation cross section to order $A_0^{2n}$ corresponds to an expansion 
up to $n^{th}$ order in opacity.

Taking the high energy limit and the soft radiation approximation ($x<<1$), one focuses on isolating 
the leading behaviour in $x$ that arises from the large number of  interference diagrams at a given 
order of opacity. As a result of the approximations made, one recovers the BDMPS condition that 
the leading behaviour in $x$ is contained solely in gluon re-scattering diagrams.  This results in the 
expression for the inclusive energy distribution for gluon radiation off an in-medium produced parton as~\cite{Wiedemann:2000tf},

\bea
\fl x\frac{dI}{dx} &=& \frac{\A_s C_R}{(2\pi)^2 x^2} 2 {\rm Re}\!\!\! \int\limits_{\zeta_0}^\infty \!\!\!d y_l  
\!\!\int\limits_{y_l}^{\infty} \!\!\! d \bar{y}_l 
\!\!\int \!\!\! d\vec{u}\!\!\!\! \int\limits_0^{\chi x p^+} \!\!\! d \vec{k} e^{- i \vk \x \vec{u} - \hf \int d\zeta n(\zeta) \sg(\vec{u}) }
\!\!\!\!\frac{\prt^2}{\prt y \prt u}\!\!\!\!\!\!\!\!\int\limits_{\vec{y}=0=\vecr(y_l)}^{\vec{u}=\vecr(\bar{y})}\!\!\!\!\! \mathcal{D}r 
e^{i\int d\zeta \frac{xp^+}{2} \left( |\dot{\vecr}|^2  - \frac{n(\zeta) \sg(\vecr) }{i xp^+}\right)}, \label{ASW2}
 \eea 
\nt
where, as always, $k_\perp$ is the transverse momentum of 
the radiated gluon and $xp^+$ is its forward momentum. 
The vectors $\vec{y}$ and $\vec{u}$ represent
 the transverse locations of the emission of the gluon in 
the amplitude and the complex conjugate whereas $y_l$ and $\bar{y}_l$
 represent the longitudinal positions.  The density of 
scatterers in the medium at location $\zeta$ is $n(\zeta)$ and
 the scattering cross section is $\sg(r)$. In this form, the 
opacity is obtained as $\int n(\zeta) d \zeta$ over the extent of the medium.

Numerical implementations of this scheme have focused on
 two separate regimes. In one case, $\sg(r)$ is replaced with a
dipole form $Cr^2$ and one solves the harmonic oscillator like
 path integral. This corresponds to the case of multiple 
soft scatterings of the hard probe. In the other extreme, one expands 
the exponent as a series in $n\sigma$; keeping 
only the leading order term corresponds to the picture of gluon 
radiation associated with a single scattering. In this second form, the analytical 
results of the ASW scheme formally approach those of the GLV reaction 
operator expansion~\cite{Wiedemann:2000tf}.  

In either 
case, the gluon emission intensity distribution has been found to 
be rather similar, once scaled with the characteristic 
frequency in each case. Beyond this, the calculation of the total 
energy loss in the ASW scheme follows a procedure 
similar to that in the GLV approach, where, a probabilistic scheme to 
lose an energy fraction $\ep$ in multiple independent 
emissions is set up, as in Eq.~\eref{GLV3}. This is used to 
calculate the medium modified fragmentation function~\cite{Eskola:2004cr}, as in 
Eq.~\eref{GLV4}.

%%%%%%%%%%%%%%%%%%%%%%%%%%%%%%%%%%%%
%%%%%%%%%%%%%%%%%%%%%%%%%%%%%%%%%%%%

\subsection{Finite temperature field theory approach}

%%%%%%%%%%%%%%%%%%%%%%%%%%%%%%%%%%%%
%%%%%%%%%%%%%%%%%%%%%%%%%%%%%%%%%%%%

In this scheme, often referred to as the Arnold-Moore-Yaffe (AMY) approach, the energy loss of hard 
jets is considered in an 
extended medium in equilibrium at asymptotically high temperature $T \ra \infty$. 
Due to asymptotic freedom,  the coupling constant 
$g \ra 0$ at such high temperatures and a power counting scheme emerges from the 
ability to identify  a hierarchy of parametrically separated  
scales $T >> gT >> g^2 T$ \etc
In this limit, it becomes possible to construct an effective field theory of soft modes, \tie, $p \sim gT$ by 
summing contributions from  hard loops with $p \sim T$, into effective propagators and vertices~\cite{Braaten:1989kk}. 

One assumes a hard on-shell parton, with energy several 
times that of the temperature, traversing such a 
medium, undergoing soft scatterings with momentum transfers $\sim gT$ off other hard 
partons in the medium. Such soft scatterings induce collinear radiation from the parton, with 
a transverse momentum of the order of $g T$. The formation time for such collinear 
radiation $\sim 1/(g^2T) $ is of the same order of magnitude as the mean free time 
between soft scatterings~\cite{Arnold:2001ba}. As a result, multiple scatterings of the 
incoming (outgoing) parton 
and the radiated gluon need to be considered to get the leading order gluon radiation rate. 
One essentially calculates the imaginary parts of infinite order ladder diagrams such as 
those shown in the right panel of Fig.~\ref{fig1}; this is done by means of integral 
equations~\cite{Arnold:2002ja}.

The imaginary parts of such ladder diagrams yield  the $1\ra2$ decay 
rates of a hard parton $(a)$ into a radiated gluon and another parton $(b)$ $\Gamma_{bg}^a$.  These 
decay rates are then used to evolve hard quark and gluon distributions from the initial hard 
collisions,  
when they are formed, to the time when they exit the medium, by means of a Fokker-Planck 
like equation~\cite{Jeon:2003gi}, which is written schematically as, 

\bea
\fl \frac{d P_a(p)}{d t} = \int dk \sum_{b,c} \left[ P_b(p+k) \frac{d\Gamma^b_{ac}(p+k,p)}{dk dt} 
- P_a(p) \frac{d \Gamma^a_{bc}(p,k)}{dk dt}  \right] . \label{AMY1}
\eea

The initial distributions are taken from a factorized hard scattering cross section as in 
Eq.~\eref{basic_cross}. A medium modified fragmentation function is modelled by the 
convolution of the vacuum fragmentation functions with the hard parton 
distributions, at exit, to produce the final hadronic spectrum~\cite{Turbide:2005fk},

\bea
\fl \tilde{D}^h(z) = \int dp_f \frac{z\p}{z} \sum_a P_{a} (p_f;p_i) D^{h}_a(z\p). \label{AMY2}
\eea

\nt
Where, the sum over $a$ is the sum over all parton species. The two momentum fractions are 
$z=p_h/p_i$ and $z\p=p_h/p_f$, where $p_i$ and $p_f$  are the momenta of the hard partons 
immediately after the hard scattering and prior to exit from the medium. The integral above 
depends on the path taken by the parton through the medium, which in turn depends on the 
location of origin of the jet and its angle with respect to the reaction plane. The model of the 
medium is essentially contained in the space-time profile chosen for the temperature. 

%%%%%%%%%%%%%%%%%%%%%%%%%%%%%%%%%%%%
%%%%%%%%%%%%%%%%%%%%%%%%%%%%%%%%%%%%
%%%%%%%%%%%%%%%%%%%%%%%%%%%%%%%%%%%%

\section{Comparisons to data and future directions}

%%%%%%%%%%%%%%%%%%%%%%%%%%%%%%%%%%%%
%%%%%%%%%%%%%%%%%%%%%%%%%%%%%%%%%%%%
%%%%%%%%%%%%%%%%%%%%%%%%%%%%%%%%%%%%

\begin{figure}
\hspace{1cm}
\resizebox{2in}{2in}{\includegraphics[1in,3in][11in,15in]{glv.eps}}
\hspace{1cm}
\resizebox{2in}{2in}{\includegraphics[0.0in,0.25in][5in,5.25in]{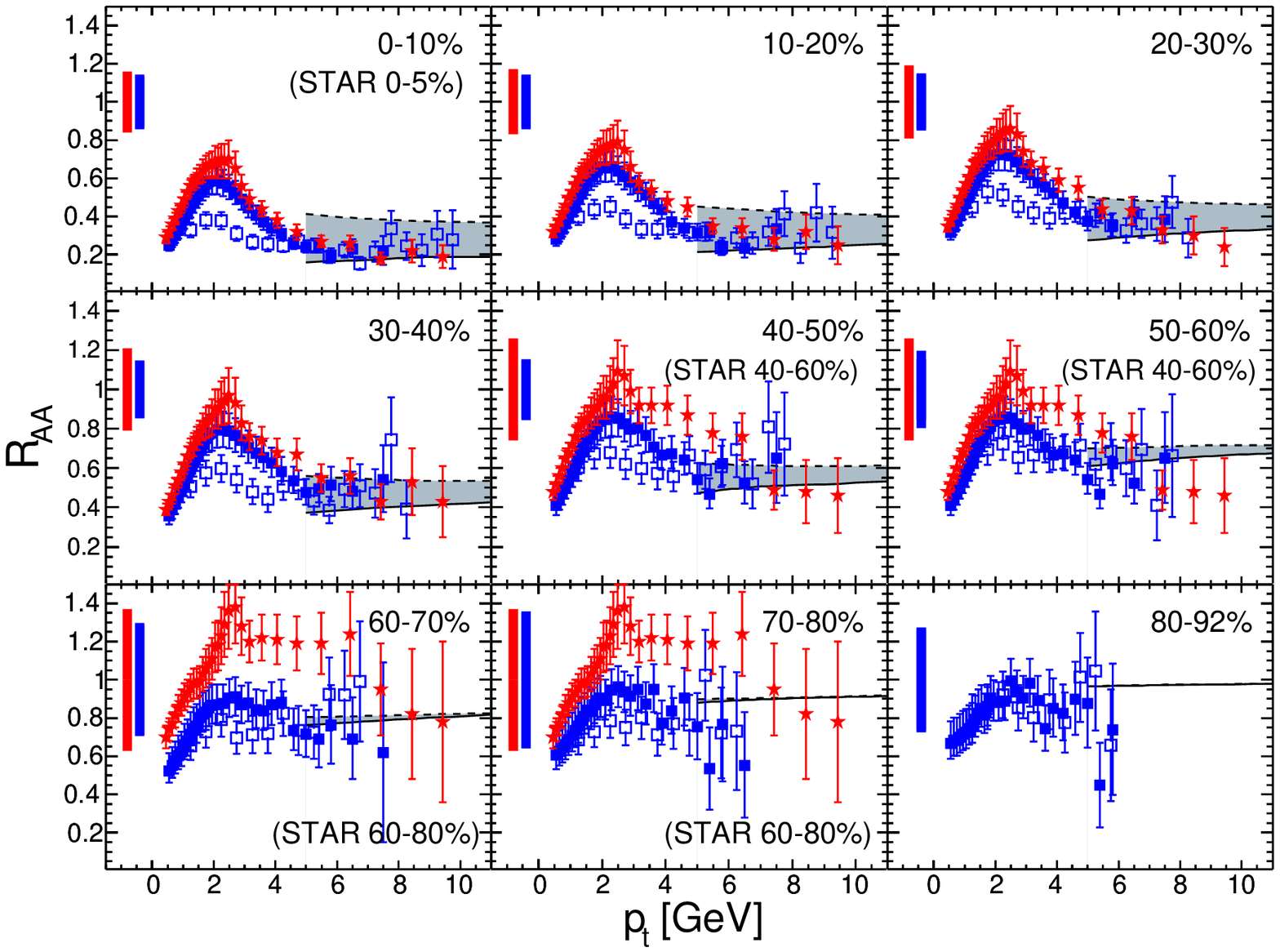}}
\caption{$R_{AA}$ as a function of $p_T$ and centrality in GLV (left) and ASW (right) 
compared to experimental data. 
The ASW plots used correspond to the PQM version of Ref.~\cite{Dainese:2004te}. The 
GLV plots are from Ref.~\cite{Vitev:2005he}.}
\label{fig2}
\end{figure}

\begin{figure}
\hspace{1cm}
\resizebox{2in}{2in}{\includegraphics[1.0in,0.5in][7in,7.9in]{amy.eps}}
\hspace{1cm}
\resizebox{2in}{2in}{\includegraphics[0.0in,2.1in][4.7in,8.7in]{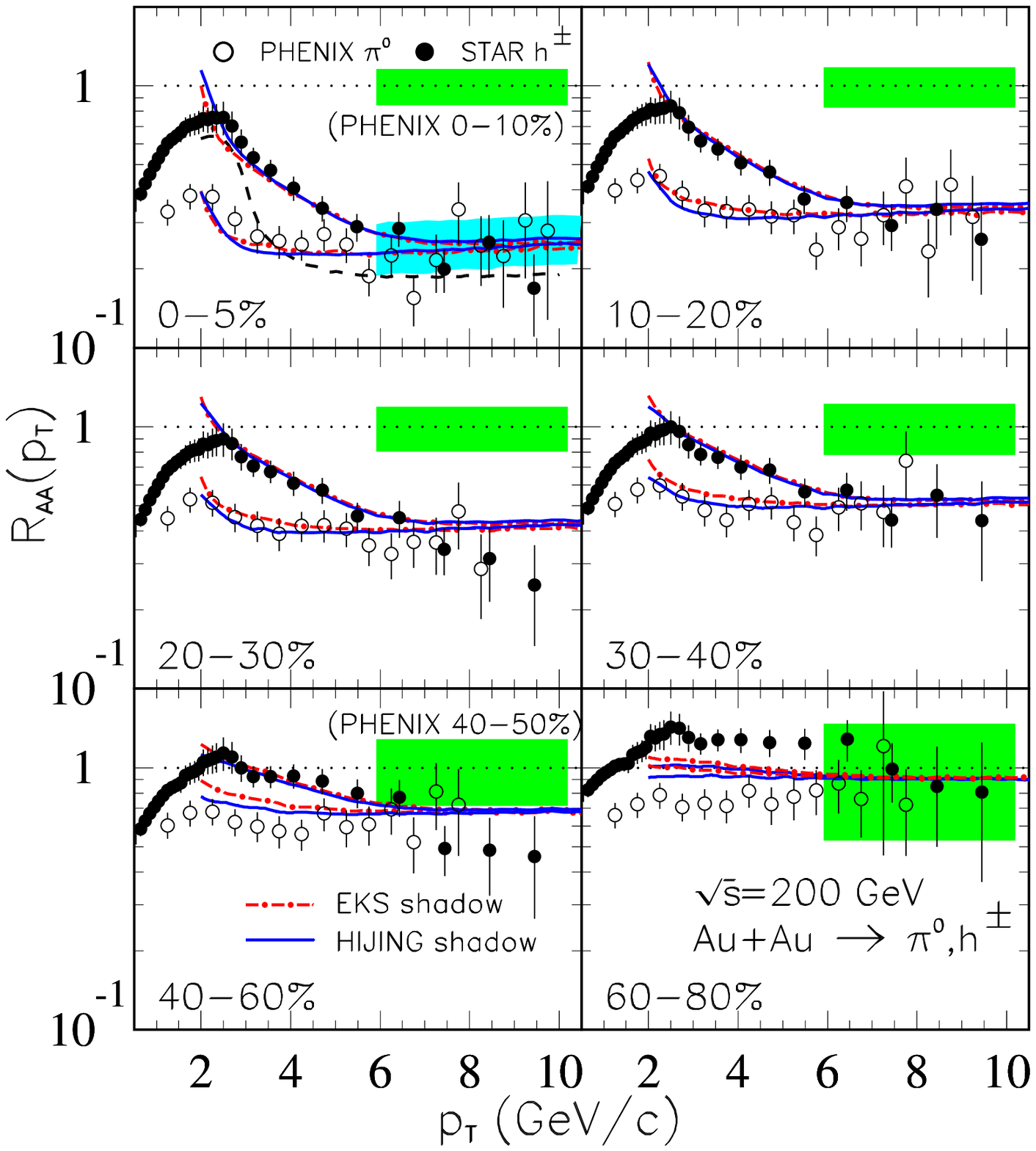}}
\caption{$R_{AA}$ as a function of $p_T$ and centrality in AMY~\cite{simon} (left panel) and HT~\cite{Wang:2003mm} 
(right panel) compared to experimental data. }
\label{fig3}
\end{figure}

At the energies of collision at RHIC, complete jets cannot be resolved from 
the large background of particles produced; energy loss is deduced from a leading 
particle analysis.
The primary observable in this regard is the nuclear modification factor 
for high $p_T$ particles as a function of $p_T$ and as a function of the 
centrality of collision. All four schemes have made successful comparisons to the 
available data as shown in Figs.~\ref{fig2},\ref{fig3}. While the qualitative 
descriptions of the medium differ between the various models, all such descriptions 
may be subjected to comparison via a single parameter: the transport coefficient 
$\hat{q}$, defined as the average squared transverse momentum imparted to the 
jet per unit path length traversed in the medium.

In all four schemes, this one parameter or 
one directly related to it is tuned to fit the data. 
In the ASW scheme, $\hat{q}$ is the fit parameter. 
In the GLV scheme, it depends on the density of scattering centers 
($dN/dy$, which is the fit parameter) and the screening 
length in the medium. In the higher twist scheme, $\hat{q}$ 
depends on the gluon density of the medium, which constitutes 
the fit parameter. In the AMY scheme, the fit 
parameter is the temperature ($T$) and 
$\hat{q}$ depends on $T$. It is in these derived 
values of $\hat{q}$ that differences arise between the 
various schemes. The GLV scheme uses an 
effective length as a substitute for a full space-time 
density profile and quotes a value of 
$\hat{q}\leq 1 \gev^2/$fm, averaged over 
the space-time profile of the collision.  The closely 
related ASW scheme quotes a value of 
$\hat{q}\leq 1 \gev^2/$fm for calculations employing an a similar 
effective length of $\sim6$ fm.
However, for realistic geometries, a value of $5$-$15\gev^2/$fm is 
deduced. The H-T 
scheme quotes a value of $\sim 1-2\gev^2/$fm as the 
maximum value at a time $\tau=1$fm, 
while the AMY approach calculates a 
$\hat{q}\sim 2 \gev^2/$fm at a 
$T\simeq300$MeV; in both cases, this implies a 
$\hat{q}\leq 1 \gev^2/$fm averaged over the life-time of the collision.

Two comments are in order: of the four schemes, the ASW and the GLV are most closely related 
in terms of basic physical picture and calculational scheme used whereas the HT and the AMY 
approaches are notably different in their origins. As a result, the difference in the 
estimates of $\hat{q}$ between the ASW and GLV approaches and the similarity between 
the AMY and the HT schemes is remarkable. A heuristic explanation for the former has 
been provided by the ASW authors: in a calculation with realistic geometry,
 the average path length 
traversed by a jet was found to be  $L\sim2$ fm, three times smaller than that used in 
the fixed length estimates; as energy loss depends quadratically on $L$, the 
effective $\hat{q}$ extracted was nine times larger.   
The second observation (regarding AMY and HT) 
may be understood as the result of the series of approximations made in 
each case. It has been demonstrated that the high temperature medium described by the HTL 
effective theory may also be understood in terms of a Vlasov theory of hard particles with mometa 
$\sim T$ moving under the influence of soft $\sim gT$ fields~\cite{Blaizot:1993zk}. 
The jet parton is identified as 
similar to the hard modes $\sim T$ in terms of its propagation through the medium.  
A similar picture of 
hard partons moving under the influence of soft fields via the effect of a colour Lorentz 
force, also arises in an all twist re-summation, if it is assumed that the colour correlation 
length in the medium is short~\cite{Fries:2002mu}.

\begin{figure}
\hspace{-0.7cm}
\resizebox{2.6in}{2in}{\includegraphics[0.0in,0.7in][7.5in,6.0in]{near_side.eps}}
\hspace{-2.3cm}
\resizebox{2.6in}{2in}{\includegraphics[-0.8in,0.4in][5.5in,7in]{double_plot_2.eps}}
\hspace{-2.3cm}
\resizebox{2.6in}{2in}{\includegraphics[-0.8in,0.3in][3.2in,2.7in]{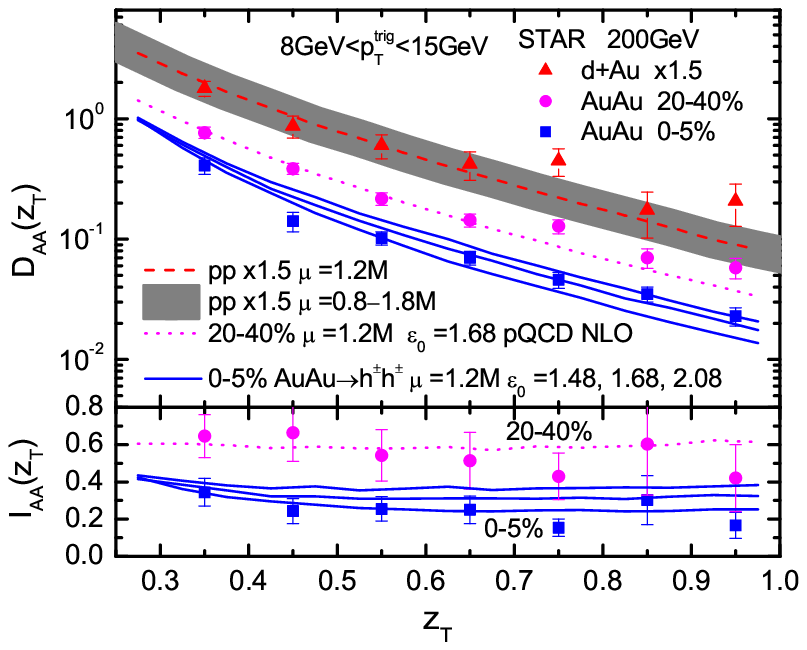}}
\caption{Left: The near side associated yield as a function of centrality of collision~\cite{Majumder:2004wh}. 
Center: the associated gluon distribution in azimuth and 
pseudo-rapidity from a hard jet just after production(Initial) and at exit from a medium 
with turbulent transverse fields (Final)~\cite{Majumder:2006wi}. 
Right: the triggered associated fragmentation function 
$D(z_T,p_T^{trig})$ and its ratio in $Au$-$Au$ with that in $p$-$p$~\cite{Zhang:2007ja}.}
\label{fig4}
\end{figure}

However, the radiation spectrum as well as the hierarchy of scales used in each of the
different approaches remains different. Single inclusive observables such as the $R_{AA}$ 
reveal an overall effect of energy loss, where the differences between the various 
schemes may be masked by a renormalization of the undetectable properties of the medium.
To demonstratively distinguish between the various schemes requires the use of differential 
probes (beyond leading hadrons) emanating from the same jet. To date, such calculations  
have been performed at the integrated level~\cite{Majumder:2004wh,Renk:2006nd}. In the 
left panel of Fig.~\ref{fig4}, calculations of the variation of the integrated yield of hadrons, 
associated with a hard trigger hadron, as a function of centrality of the collision are plotted. The yield, naturally 
depends on the detected flavour content, which depends on the ability of the detector to account for  
decay corrections. The theoretical curves represent two extreme possibilities, where 
only charged pions are detected or all charged hadrons are detected including contributions from 
all decay corrections. The slight rise with centrality in the theoretical curves is due to the larger amount 
of near side energy loss in more central collisions. 
Yet another method to test the scheme dependence of energy loss is via single inclusive probes, 
subjected to a more differential analysis \eg, the $R_{AA}$ versus the reaction plane~\cite{Majumder:2006we}. 
Such measurements will not only resolve the differences in medium profiles between the various schemes, but 
produce more detailed probes of the evolving medium as seen at the scale of the jet. 

The study of energy loss in dense matter encompasses, not merely a study of a single 
medium property $\hat{q}$ in terms of its effect on parton propagation, but through 
hard-soft correlations, can be extended to a study of how the lost energy is redistributed 
by the medium. Phenomenological studies of hard-soft correlations indicate that the 
lost energy seems to excite collective density excitations in the medium~\cite{Casalderrey-Solana:2006sq}.
However, alternate explanations in terms of Cherenkov radiation~\cite{Koch:2005sx} or 
in-medium Sudakov effects in jet radiation~\cite{Polosa:2006hb} have not 
been ruled out. 
While such a study lies outside the reach of perturbative QCD, a precise 
determination of the mechanism of energy loss and its distribution along the path of the 
jet become crucial to all model studies of energy redistribution. The requirement of 
unambiguous results, necessarily confines the 
study of energy loss mechanisms and associated measurements beyond the reach of 
final state bulk-medium  effects such as recombination, \tie, $p_T \geq 7\gev$. 
Such detailed studies will ultimately require higher statistics in the experimental measurements 
at high transverse momenta and predictions from all four schemes for the same measurement.

In spite of its long history, 
developments in the basic theory of jet modification continue to be made.
It has recently been proposed that $\hat{q}$ may posses a tensorial 
structure~\cite{Majumder:2006wi}, where the scalar 
$\hat{q} = \kd^{ij} \hat{q}_{ij}$. An example where such a situation may arise is in the 
presence of  large 
turbulent colour fields, which may be generated in the early 
plasma due to anisotropic parton distributions~\cite{weibel,Asakawa:2006tc}. These 
large fields, transverse to the beam, tend to deflect radiated gluons 
from a transversely traveling jet, preferentially, in the longitudinal directions. 
While such effects influence the solid angle distributions of the radiated gluons 
around the originating parton, they do not have a considerable effect on the 
total energy lost. 
Such phenomena may yield an explanation for the ridge like structure seen in the 
near side correlations. In a Ref.~\cite{Majumder:2006wi}, 
a quantitative estimate of this broadening was made and the results were found to 
decrease with increasing energy of the radiated gluon as shown in the center panel of 
Fig.~\ref{fig4}. Here the angular profile of radiated gluons in azimuth ($\phi$) and 
pseudo-rapidity $(\eta)$ at production and at exit from an expanding medium 
is plotted. Results are for two energies of the original jet of 10 and 
20 GeV, formed at a depth of 3 fm inside the medium. The radiated gluon always 
carries a fraction $x=0.4$ of the jet's energy. 

In most jet quenching calculations, the hard cross section 
for both single inclusive and double inclusive observables has to date been computed 
at leading order~\cite{Renk:2006nd,Qiu:2003pm}. 
Recently, such calculations have been 
extended to next-to-leading order (NLO)~\cite{Zhang:2007ja}. Here, the authors 
computed both single inclusive as well as triggered  back-to-back distributions of 
leading hadrons. 
The right most panel of Fig.~\ref{fig4} shows the $p_T^{trig}$
weighted differential cross section to produce an associated hadron with a given $p_T^{assoc}\!\!$, given a 
trigger hadron [also referred to as the triggered fragmentation function $D(z_T, p_T^{trig.})$, where $z_T=p_T^{assoc}/p_T^{trig}$]. Also plotted is 
the ratio of this cross section in $Au$-$Au$ with that in $p$-$p$ referred to as the $I_{AA}$.
Such developments along with emerging computations, of heavy quark energy loss and photon 
hadron correlations hold promise for the central role of jet-medium interactions 
in heavy-ion physics.

The author thanks N.~Armesto, S.~A.~Bass, C.~Gale, S.~Jeon, C.~Loizides, G.~Moore, B.~Muller, 
T.~Renk, C.~Salgado, S.~Turbide, I.~Vitev, X.~N.~Wang and U.~Wiedemann for helpful discussions. 
Work supported in part by grants from the U.S. department of energy (DE-FG02-05ER41367) and 
National Science Foundation (NSF-INT-03-35392).

% \section*{References}

\end{document}